\documentclass[reprint,preprintnumbers,aps,prd,tightenlines,longbibliography,nofootinbib]{revtex4-2}

\usepackage{graphicx}
\usepackage{xcolor}
\usepackage[sort&compress]{natbib}
\usepackage{amsmath,amssymb,bm,bbm,slashed,subdepth,amsfonts}
\usepackage{booktabs}  %
\usepackage{xr-hyper}
\usepackage[colorlinks=true,
urlcolor=blue,
anchorcolor=blue,
citecolor=blue,
filecolor=blue,
linkcolor=red,
menucolor=blue,
linktocpage=true,
pdfproducer=medialab,
pdfa=true
]{hyperref}
\usepackage{cleveref}
\usepackage{enumerate,comment}
\usepackage{epsfig, subfigure}
\usepackage{setspace}
\usepackage{booktabs, tabularx}
\usepackage{units}
\usepackage{placeins}
\usepackage{multirow}
\usepackage{mathtools}
\usepackage[normalem]{ulem}

\graphicspath{{fig/}}

\begin{document}
\title{Identifying Monochromatic Signals in LISA and Taiji via Spectral Split: \\ Gravitational Waves versus Ultralight Dark Matter}
\author{Yue-Hui Yao$^{a,d}$} 
\author{Tingyuan Jiang$^{a,d}$}
\author{Wenyan Ren$^{b,c}$}
\author{Di Chen$^{a,c}$}
\author{Yong Tang$^{a,b,e}$}
\author{Yu-Feng Zhou$^{b,c,d}$}
\affiliation{\begin{footnotesize}
		${}^a$University of Chinese Academy of Sciences (UCAS), Beijing 100049, China\\
		${}^b$School of Fundamental Physics and Mathematical Sciences, \\
		Hangzhou Institute for Advanced Study, UCAS, Hangzhou 310024, China\\
		${}^c$Institute of Theoretical Physics, Chinese Academy of Sciences, Beijing 100190, China \\
        ${}^d$International Center for Theoretical Physics Asia-Pacific, Beijing, China \\
        ${}^e$School of Astronomy and Space Science, UCAS, Beijing 100049, China
\end{footnotesize}
}
\date{\today}

\begin{abstract}
The detection of gravitational waves (GWs) has opened a new window to explore the dark Universe. Ultralight dark matter (ULDM), an attractive candidate for dark matter, might induce monochromatic signals in gravitational-wave (GW) laser interferometers. However it is not clear how such signals are disentangled from the GWs emitted by galactic compact binaries. Here we initiate the investigation on the spectral split of monochromatic signals caused by detector's heliocentric motion in space and show the annual modulation can induce distinct structures in the spectral harmonics for GWs and ULDM, which would enable to clearly identify the nature of the signal. We show the physical parameters can be inferred with high precision using the Fisher matrix formalism. Our results provide a practical algorithm for probing ULDM and broaden the scientific objectives of future GW detectors in space, such as LISA and Taiji.
\end{abstract}

\maketitle
\clearpage

%------------------------------------------------------------
\section{Introduction} \label{sec:intro}
While overwhelming evidence points to its existence, the nature of dark matter (DM) remains elusive~\cite{cirelli2024darkmatter}. 
Among various scenarios, an intriguing one is that DM consists of ultralight bosonic particles~\cite{Chadha-Day:2021szb, PhysRevLett.38.1440, Weinberg1978, PhysRevLett.40.279, PRESKILL1983127, Dine:1982ah, Damour:1994ya, Damour:1994zq, Capozziello:2011et, PhysRevD.84.103501, PhysRevD.93.103520, Ema:2019yrd}.
These ultralight dark matter~(ULDM) particles have a large occupation number within a de Broglie volume, and their collective behavior can be well described by classical fields~\cite{PhysRevD.95.043541, annurev:/content/journals/10.1146/annurev-astro-120920-010024}.
Owing to its wave nature, ULDM exhibits rich phenomenology~\cite{Schive:2014dra, Centers:2019dyn, Vermeulen:2021epa, Chen:2021lvo, Gavilan-Martin:2024nlo, Tsai:2021lly, An:2023wij, Jiang:2023jhl, Du:2022trq, PhysRevD.111.042008, Khmelnitsky:2013lxt, Liu:2021zlt} and can alleviate the tensions of the standard cosmological model on small scales~\cite{Bullock:2017xww, deBlok:2009sp, Boylan-Kolchin:2011qkt, Tulin:2017ara}.

Equipped with cutting-edge technology, gravitational-wave~(GW) laser interferometers are among the most sensitive and powerful experimental instruments ever built, whose successful operation allowing us to ``hear" the Universe for the first time~\cite{LIGOScientific:2016aoc}. 
While these laser interferometers are primarily designed to detect gravitational waves~(GWs), 
their exquisite sensitivity to variations in optical path difference also makes them highly sensitive to ULDM.
The ULDM field oscillates at its Compton frequency and, through its interaction with the standard model~(SM) particles, exerts forces on the test masses or modulates laser phase, which changes optical path and leads to potentially detectable signals in interferometers.
Powerful constraints and projected sensitivities on the coupling strengths between the ULDM and SM particles have been obtained in this way~\cite{PhysRevLett.121.061102, PhysRevD.100.123512, PhysRevResearch.1.033187, Guo:2019ker, Vermeulen:2021epa, PhysRevD.107.063015, PhysRevD.108.083007, PhysRevD.108.095054, Kim:2023pkx, PhysRevD.109.095012, PhysRevD.110.023025, PhysRevD.110.095015, Gue:2024txz, PhysRevD.111.055031, yao2025axionlikedarkmattersearch, zhang2025probingspin2ultralightdark, miller2025gravitationalwaveprobesparticle}.

% The similarity between monochromatic GW and ULDM signals.
Due to the velocity dispersion of DM particles, the field has a finite coherence time.
When the observation time is shorter than this, the induced signals are monochromatic, with frequencies at integer multiples of the Compton frequency~\cite{PhysRevD.100.123512, Guo:2019ker, Kim:2023pkx, PhysRevD.107.063015, Amaral:2024tjg, PhysRevD.110.095015, Gue:2024txz, PhysRevD.111.055031}.
However, monochromatic signals in a GW interferometer can also originate from GWs from compact binary inspirals or instrument artifacts. 
Therefore, when such a signal is detected, it is crucial to determine its physical nature. In Refs.~\cite{PhysRevD.108.083007, xu2025distinguishingmonochromaticsignalslisa}, we proposed to distinguish these signals in space-based GW detectors~\cite{amaroseoane2017laser, Hu:2017mde, Luo_2016} by utilizing the different responses of various interferometry channels to GWs and ULDM. However, the sensitivities of such channels are also affected.

\begin{figure*}
    \begin{center}
    \includegraphics[width=0.9\linewidth]{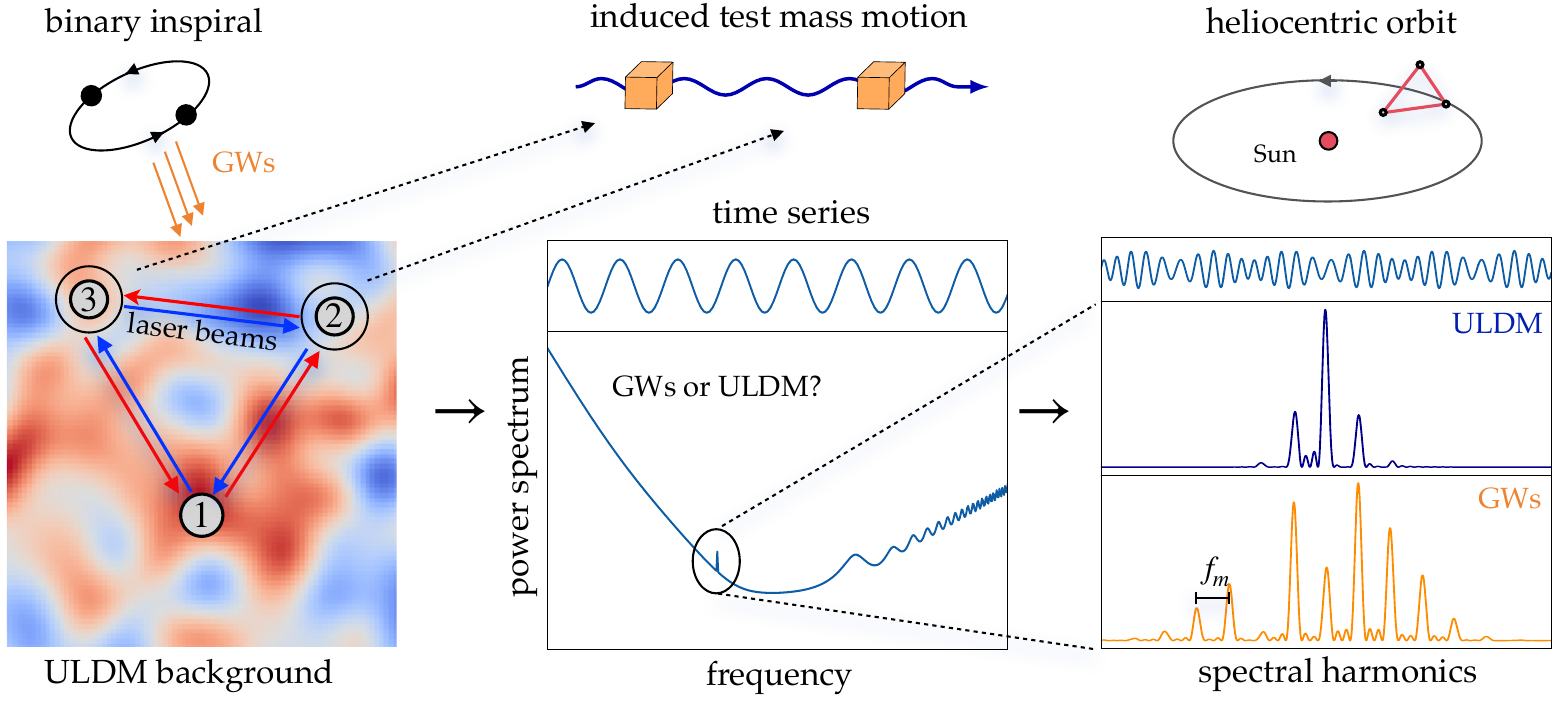}
    \end{center}
    \caption{Schematic of the spectral splitting induced by the detector's heliocentric motion. 
    ULDM exerts oscillatory forces on the test masses inside the spacecrafts, altering their inter-spacecraft distances and producing a monochromatic signal similar to that induced by GWs from binary inspirals. The detector's motion splits a monochromatic signal into a series of harmonics separated by the orbital frequency $f_m = 1 / \textrm{yr} \approx 3 \times 10^{-8}~\textrm{Hz}$. Due to their distinct couplings to the detector and kinetic properties, the modulated signal spectra of GWs and ULDM exhibit distinct features, which offers a way to distinguish them.}
    \label{fig:schematic}
\end{figure*}

Here we present a new approach which leverages the modulated signal spectra arising from the detector's heliocentric motion (Appendix~\ref{ap:orbit}).  
% Explain modulation
As illustrated in Fig.~\ref{fig:schematic},
the periodic motion of detector modulates the signal, transforming an originally monochromatic signal into a series of harmonics spaced by the orbital frequency in the Fourier domain.
We demonstrate that due to their distinct couplings to the detector and kinetic properties, 
the modulated signal spectra of GWs and ULDM exhibit distinct features.
Notably, the GW spectrum includes significant higher-order harmonics than the ULDM spectrum.
Therefore, observing the corresponding spectral pattern directly reveal the signal origin.
Using the Fisher matrix formalism, we also examine how well signal parameters can be extracted from observations.
Our results show that the parameters of ULDM, such as its mass, can be determined with high precision using interferometers.

%------------------------------------------------------------
\section{Detector Response}~\label{sec:framework}
A space-based GW laser interferometer, such as LISA~\cite{amaroseoane2017laser} and Taiji~\cite{Hu:2017mde}, typically consists of three heliocentric spacecrafts, which form a quasi-equilateral triangle constellation.
%with a few million kilometers arm length in space. 
Each spacecraft exchanges laser beams with the other two, monitoring the physical distances between the free-falling test masses in different spacecraft.
Variations in inter-spacecraft distances effectively cause fluctuations in laser frequency. We denote the recorded relative laser frequency fluctuations in the six links by $y_{rs}(t)$, where $r,s = 1,2,3$ labels the spacecraft.

\textbf{GWs:} The passing GWs with propagation direction $\hat{k}$ perturb the inter-spacecraft distance, and contribute to $y_{rs}(t)$ in the link with spacecraft $r$ receiving and spacecraft $s$ sending by~\cite{Romano:2016dpx, babak2021lisasensitivitysnrcalculations, Romano_2024}
%\footnote{Throughout the paper we use natural units $c=\hbar=1$.}
\begin{equation} \label{eq:1link GW signal}
    y_{rs}(t) \supset \hat{n}_{rs}:\frac{\boldsymbol{h}(t_s - \hat{k}\cdot\mathbf{x}_s) - \boldsymbol{h}(t - \hat{k}\cdot\mathbf{x}_r)}{2(1-\hat{k}\cdot\hat{n}_{rs})}:\hat{n}_{rs}, 
\end{equation}
where ``:" denotes tensor contraction, 
$\mathbf{x}_r\equiv \mathbf{x}_r(t)$ and $\mathbf{x}_s\equiv \mathbf{x}_s(t_s)$ are the spacecrafts' time-dependent position vectors, and the unit vector
$\hat{n}_{rs} = (\mathbf{x}_r-\mathbf{x}_s)/|\mathbf{x}_r-\mathbf{x}_s|$. Here
$t_s$ is the time at which the laser beam leaves the spacecraft $s$ and is obtained by solving the transcendental equation $t-t_s = |\mathbf{x}_r(t)-\mathbf{x}_s(t_s)|$.
Here $\boldsymbol{h}(\xi)$ denotes the GW tensor.
For monochromatic sources, it takes the form $\boldsymbol{h}(\xi) = e^{i2\pi f \xi}\sum_{p} H_p\mathbf{e}^p(\hat{k})$, where $f$ is frequency and $H_p$ is the amplitude of mode $p$ defined relative to a set of polarization basis $\mathbf{e}^p(\hat{k})$ (see Appendix~\ref{ap:GW}).

\textbf{ULDM:} DM particles follow a velocity distribution in our galaxy~\cite{Freese:2012xd}, and we can model ULDM as a superposition of plane waves, each one corresponding to a collection of DM particles occupying the same region in momentum space~\cite{PhysRevA.97.042506, PhysRevD.97.123006, PhysRevD.111.015028, Kim:2023pkx, PhysRevD.110.095015}. Effectively a vector ULDM in the vicinity of solar system takes the form
\begin{equation} \label{eq:DM mono}
    \mathbf{A}(x) = e^{imt}\sum_p a_p(x)  \hat{x}^p ,
\end{equation}
where $x=(t,\mathbf{x})$, $m$ is the mass of ULDM, and $a_p$ are complex amplitudes defined with respect to a set of basis vectors $\hat{x}^p$ (Appendix~\ref{ap:ULDM}). 
For well-virialized DM, the three amplitudes $a_p(x)$ are uncorrelated and vary stochastically on the scales specified by the coherence time $\tau_c=2\pi/m\sigma^2$ and coherence length $\lambda_c = \sigma\tau_c$, where $\sigma \sim 10^{-3}$ is the velocity dispersion of DM.
For $m = 10^{-17}~\text{eV}$, we have the Compton frequency $f_c = m/2\pi \approx 2.4~\textrm{mHz}$, $\tau_c \approx 4.13 \times 10^8~\text{s}$, and $\lambda_c \approx 1.24 \times 10^{11}~\text{km}$.

If the vector ULDM couples to test masses through either the baryon number or baryon minus lepton number, its oscillations induce motion of test masses, and the resulting laser frequency fluctuation is given by~\cite{PhysRevLett.121.061102, PhysRevD.108.083007}
\begin{equation} \label{eq:1link DM signal}
    y_{rs}(t) \supset g\hat{n}_{rs} \cdot
    \left[\mathbf{A}(t,\mathbf{x}_r) - \mathbf{A}(t_s,\mathbf{x}_s)\right] ,
\end{equation}
where $g$ encapsulates experiment-specific parameters as well as the coupling strength between ULDM and SM particles. For scalar ULDM $\Phi$ coupled via the trace of test mass's energy-momentum tensor~\cite{PhysRevD.110.095015, PhysRevD.111.015028}, the signal is given by Eq.~(\ref{eq:1link DM signal}) with the substitution $\mathbf{A}\rightarrow \nabla\Phi$.

The single-link data streams are further compiled into laser-noise-free data channels, using a post-processing algorithm known as time-delay interferometry~\cite{PhysRevD.62.042002, PhysRevD.65.102002, PhysRevD.72.042003, Tinto:2020fcc}.
Here, we use the standard Michelson channel $X$,
\begin{equation} \label{eq:def TDI X}
    \begin{split}
        X(t) = & \left(y_{13} + y_{31,2} + y_{12,22} + y_{21,322}\right) \\
        & - \left(y_{12} + y_{21,3} + y_{13,33} + y_{31,233}\right), 
    \end{split}
\end{equation} 
where the time-delayed series $y_{rs,k\dots l}(t) = y_{rs}(t - L_k -\cdots - L_l)$ with $L_k$ the length of the arm opposite spacecraft $k$. $X(t)$ will be the main quantity we shall analyze in both time and frequency domains.

\section{Spectral Split by Modulation}\label{sec:spectra}
Using Eqs.~(\ref{eq:1link GW signal}) and (\ref{eq:def TDI X}), 
we derive the signal induced by a monochromatic GW with a frequency $f$ as a collection of evenly-separated harmonics (Appendix~\ref{ap:GW})
\begin{equation} \label{eq:X GW modul}
\begin{split}
    X(t) 
    &\simeq 16\pi^2f^2L^2 \sum_{n=-\infty}^{+\infty} e^{i2\pi (f+ nf_m)t} \\
    &\qquad\times\sum_{l=-2j}^{2j} \mathcal{D}^{(n-l)}(f,\hat{k}) \sum_{p=+,\times} G^{(l)}_p(\hat{k})H_p ,
\end{split}
\end{equation}
where $f_m = 1/\textrm{yr}$ is the orbital frequency, $L$ is the detector's fiducial arm length,
and $jf_m$ is the frequency associated with the rotation of $\hat{n}_{rs}$. For the analytical Keplerian orbital model we considered, $j=2$. The coefficients $G^{(n)}_p$ and $\mathcal{D}^{(n)}$ are related to the orbital modulation.

Orbital modulation arises from two respects.
First, the time-varying orientation of the detector, captured by the geometric coefficients $G_p^{(n)}(\hat{k})$, which depend on the source's sky location.
Their explicit forms are provided in Table~\ref{tab:GW harm cof} for the considered orbital model.
Second, there is relative motion between the detector and GW sources, and the detector experiences different GW phases during a orbital cycle.
This effect is represented by the Doppler coefficients $\mathcal{D}^{(n)}(f,\hat{k})$,
and their magnitudes are governed by $\epsilon=2\pi fr_{\odot}\sin\theta$, which quantifies the phase variation experienced by the detector. Here, $\theta$ is polar angle of the GW source, and $r_{\odot} \approx1\textrm{AU}$ is the distance between the constellation's guiding center and the Sun.
The infinite series in Eq.~(\ref{eq:X GW modul}) can be truncated at $|n-l|\sim \epsilon$. 
For GWs with frequency $f \sim 0.1$mHz, $\epsilon \lesssim 0.3$, and only the first few harmonics are important.
For $f \sim 10$mHz, $\epsilon \lesssim 30$, and a substantial fraction of signal power is distributed across higher-order harmonics.

\begin{figure}
    \includegraphics[scale=0.9]{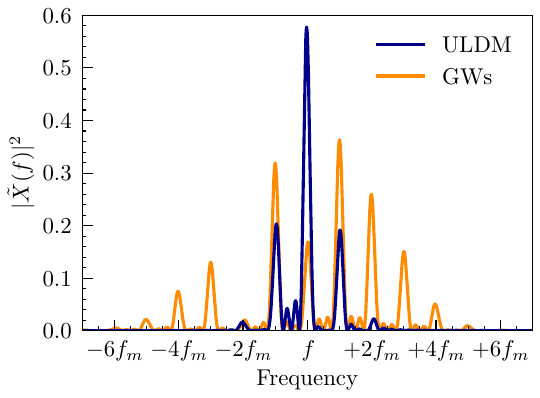}
    \caption{The modulated spectra in the analytical Keplerian orbital model. 
    We adopt the following parameter values in the plot: both the GW frequency and ULDM's Compton frequency are set to $1~\textrm{mHz}$; the GW amplitudes are $H_+ = 1$, $H_{\times} = i$; the ULDM amplitudes are $a_x = 0$, $a_y = 1$, $a_z = i$. 
    The GW source is located at the polar angle $\theta=\pi/3$ and azimuth angle $\phi=0$.
    We assume a 4-year observation time and $L = 2.5 \times 10^6~\textrm{km}$ with the initial orbital parameters $\kappa = \lambda = 0$. 
    The spectra are normalized to the peak amplitude of the corresponding unmodulated monochromatic spectrum.
    }
    \label{fig:dm2gw_X}
\end{figure}

The GW spectrum $|\tilde{X}(f)|^2$ in frequency domain is obtained by Fourier transforming Eq.~(\ref{eq:X GW modul}).
In Fig.~\ref{fig:dm2gw_X}, we plot $|\tilde{X}(f)|^2$ for a specific set of source parameters, normalized to the peak of the corresponding monochromatic spectrum without modulation.
The spectrum consists of a series of harmonics separated by the orbital frequency $f_m$ and can be divided into two parts. 
The first part includes harmonics up to $n = \pm2j$, arising from both the detector's time-varying orientation and its orbital motion.
The second part contains harmonics beyond $n = \pm2j$, which originate solely from the detector's motion.
The sidebands adjacent to each harmonics are due to the finite duration of signal  (Appendix~\ref{ap:finite time}).
In general, due to the phase difference between the two complex polarization amplitudes, the spectrum is asymmetric.

Using Eqs. (\ref{eq:1link DM signal}), (\ref{eq:def TDI X}), and (\ref{eq:DM mono}),  
we find that the ULDM signal consists of velocity-independent and velocity-dependent terms, which correspond to the contributions from the field itself and its gradient, respectively.
However for the mass range of interest, the velocity-dependent terms are suppressed relative to the velocity-independent terms due to the nonrelativistic nature of DM (Appendix~\ref{ap:ULDM}). 
Keeping only the velocity-independent terms, the ULDM signal is given by
\begin{equation} \label{eq:X DM modul}
    X(t) \simeq -
    2im^3L^3 \sum_{n=-j}^{j} e^{i2\pi (f_c+ nf_m)t} \sum_{p} \mathcal{G}^{(n)}_p a_p(\mathbf{x}_1),
\end{equation}
where $a_p(\mathbf{x}_1)$ are the amplitudes at the spacecraft 1's position.
We do not indicate the time dependence of the amplitudes, as their intrinsic evolution can be neglected during an observation of several years for ULDM with mass $\sim10^{-17}~\textrm{eV}$.
The summation over $n$ is done for integers from $-2$ to $2$.
Analogous to the GW case, the geometric coefficients $\mathcal{G}^{(n)}_p$ represent the modulation due to the detector's time-varying orientation. The index $n$ lies in $\pm1j$, since the coupling in Eq.~(\ref{eq:1link DM signal}) is linear.
% \footnote{This is not the case for tensor ULDM. Nevertheless, its nonrelativistic nature still leads to a spectrum that is different from that of GWs.}
However, there is an important distinction.
While the amplitudes $a_p(\mathbf{x}_1)$ depend on time through the spacecraft's position, the spatial scale of the orbit is much smaller than the field coherence length for the mass range of interest. The change in amplitude is estimated to be 
$|\Delta a_p/a_p| \sim 2\pi r_{\odot}/\lambda_c\approx 6 \times 10^{-3} \ll 1 $ for $m\sim10^{-17}~\text{eV}$, which is negligible. Thus, the amplitudes can be treated as constant, and Doppler modulation is absent in the ULDM case.

In Fig.~\ref{fig:dm2gw_X}, we also show the modulated ULDM spectrum for comparison, which exhibits fewer harmonics compared to GWs.
This has important implications for detection.
% matching GW signal with ULDM templates
Suppose the data contains a GW signal and we perform matched filtering using ULDM models, there will always be templates that yield a non-negligible signal-to-noise ratio~(SNR) by capturing the few dominant harmonics of the signal, e.g. the $n=(-1,0,1)$ or $n=(0,1,2)$ GW harmonics shown in Fig.~\ref{fig:dm2gw_X}.
However, because the GW spectrum includes higher-order harmonics not present in the ULDM case, matched filtering with the correct GW template extracts the full signal power and yields a higher SNR than that obtained with a ULDM template.
For the GW parameters used in Fig.~\ref{fig:dm2gw_X} and an SNR of $10$, the SNR obtained using the ULDM templates is less than $5.4$.
% matching ULDM signal with GW templates
On the other hand, if data contains a ULDM signal, some GW templates may also yield SNRs close to that obtained with ULDM, as they match the signal through a subset of their spectral content.
However, since GW templates include high-order harmonics, their absence in the observed data can only be attributed to noise fluctuations, thereby reducing the likelihood of GW templates.

Above we have derived the modulated spectra in $X$ channel analytically using the rigid adiabatic approximation~\cite{PhysRevD.69.082003}. In Appendix~\ref{ap:simulation}, we obtain the modulated spectra through numerical simulations without the approximation and find good agreement between the analytical and numerical results.

%------------------------------------------------------------
\section{Parameter estimation} \label{sec:pe}
Next we examine how well we can measure signal parameters using the Fisher matrix formalism. 
In the high SNR limit and assuming uniform priors for the signal parameters $\boldsymbol{\lambda} = \{\lambda_1,\lambda_2,\cdots\}$, the covariance of the posterior probability distribution are given by the inverse of the Fisher matrix~\cite{PhysRevD.77.042001}: 
\begin{equation}
    \Gamma_{ij} = \left(\frac{\partial h}{\partial \lambda_i} \,\middle|\, \frac{\partial h}{\partial \lambda_j}\right) ,
\end{equation}
where $h$ denotes the signal, and the noise-weighted inner product is defined by
\begin{equation}
    (d_1\mid d_2)=4\, \mathrm{Re}\int_0^\infty df \frac{\tilde{d}_1^*(f)\tilde{d}_2(f)}{S_n(f)} ,
\end{equation}
where $S_n(f)$ is the one-sided power spectral density of the detector noise.

\begin{figure*}[htbp]
    \centering
    \subfigure[GW]{
        \includegraphics[width=0.47\textwidth]{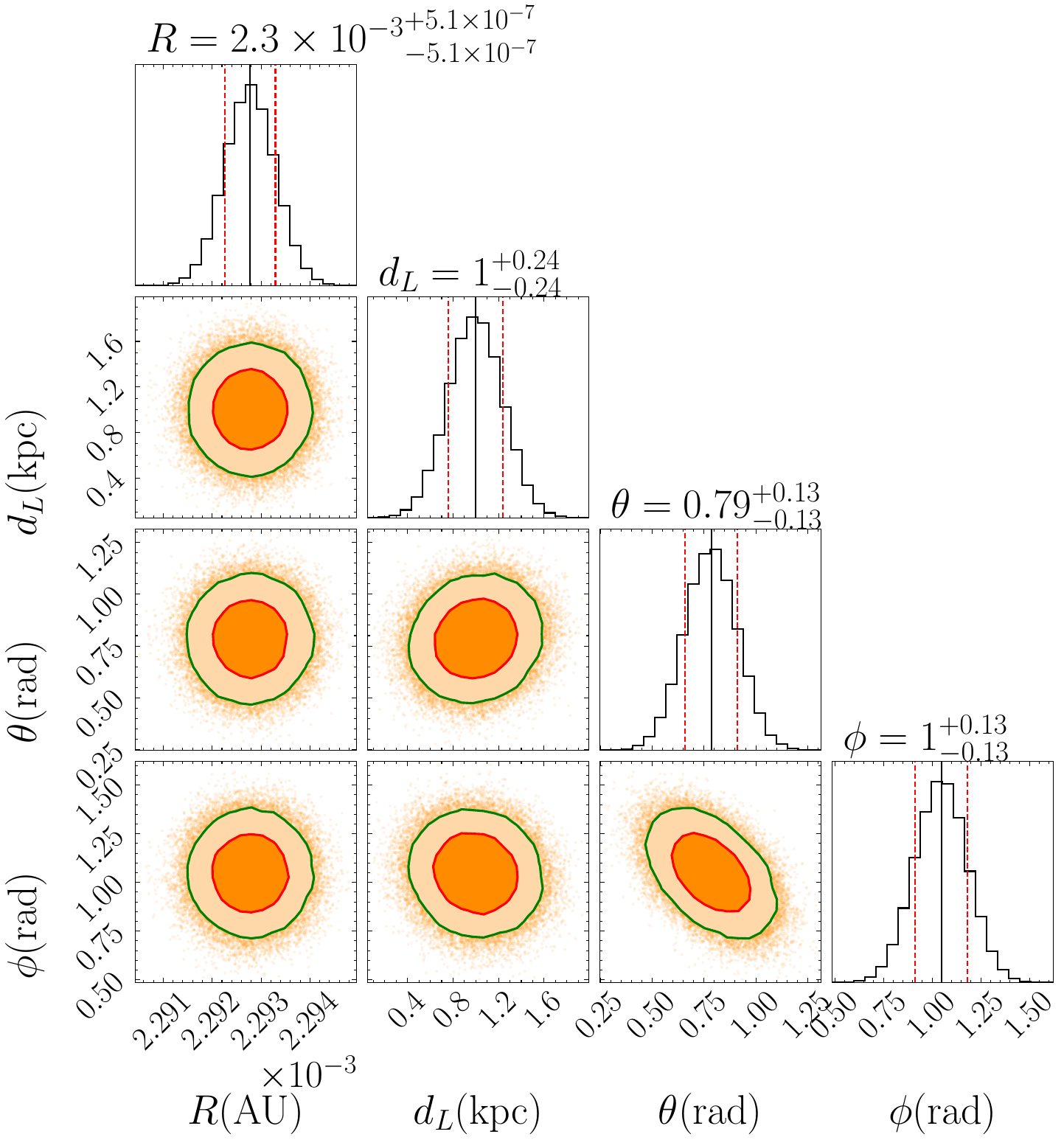}
        \label{fig:Fisher_GW}
    }%
    \hspace{0.01\textwidth} % 控制左右间距
    \subfigure[DM]{
        \includegraphics[width=0.47\textwidth]{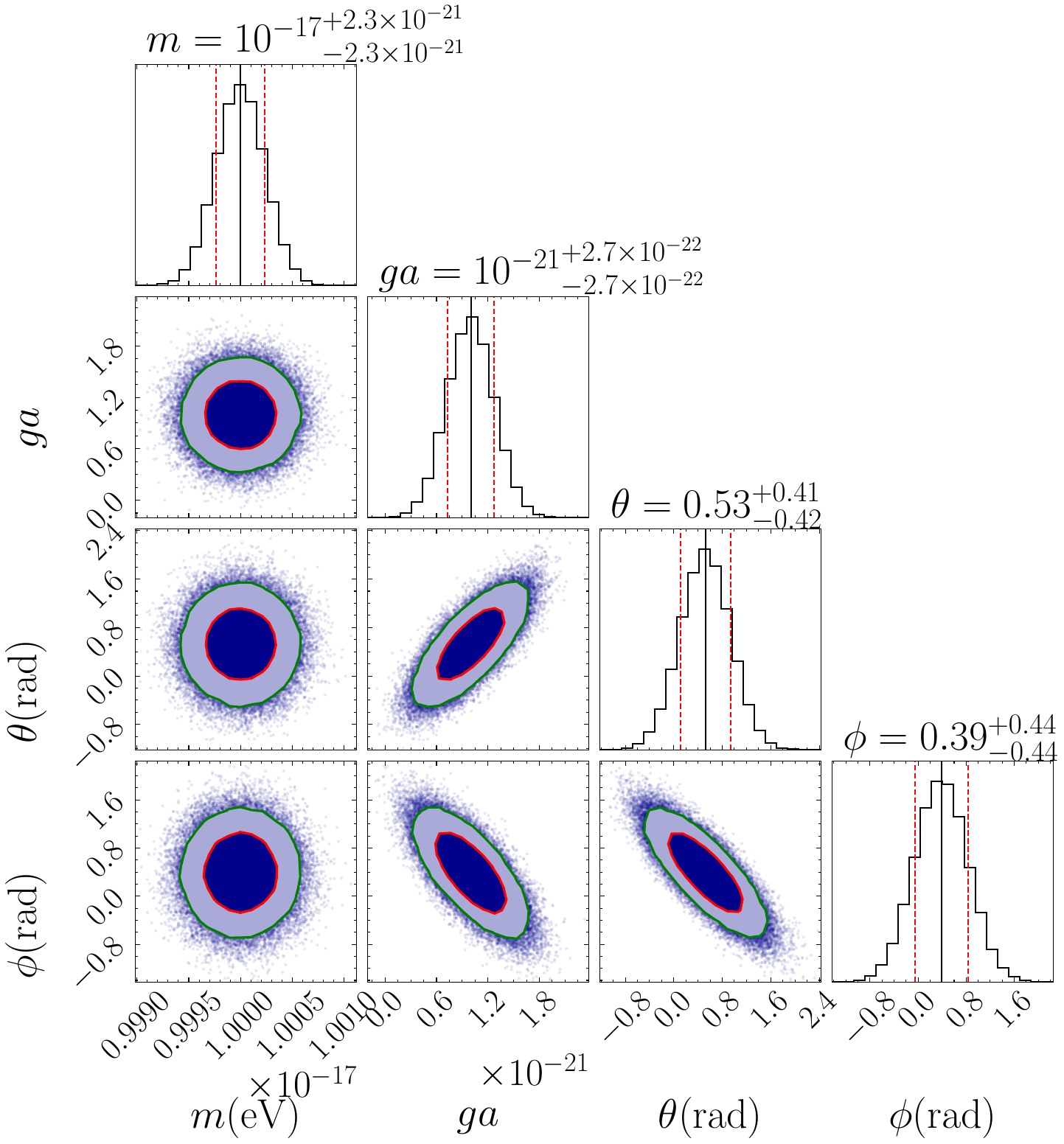}
        \label{fig:Fisher_DM}
    }
    \caption{(a) The posterior distributions for a binary system with $\mathrm{SNR}\approx 4.2$, assuming a four-year observation and the projected noise performance of LISA.
    The source parameters are: component masses $m_1=1~M_{\odot}$ and $m_2=2~M_{\odot}$, $R=2.3\times10^{-3}~\mathrm{AU}$ (corresponds to a GW frequency of $1~\textrm{mHz}$), $d_L=1~\mathrm{kpc}$, $\theta=\pi/4$, $\phi=\pi/3$, $\iota=\pi/4$, $\psi=0$, and $\varphi_0=0$.
    Red and green contours represent the $1\sigma$ and $2\sigma$ confidence regions, respectively.
    The $1\sigma$ uncertainties are: $\sigma_{R}/R \approx 10^{-4}$, $\sigma_{d_L}/d_L \approx 0.24$, $\sigma_{\theta}/\theta \approx 0.16$, and $\sigma_{\phi}/\phi \approx 0.13$. (b) The posterior distributions for a ULDM field realization with $\mathrm{SNR}\approx 6.21$, assuming a four-year observation and the projected noise performance of LISA. The parameters are: $m=10^{-17}~\mathrm{eV}$ (corresponds to a Compton frequency of $f_{c} = 2.4~\mathrm{mHz}$), $ ga=10^{-21}$, $\theta=\pi/6$, $\phi=\pi/8$, $e=0.72$, $\psi=0$, and $\varphi_0=0$.
    Red and green contours represent the $1\sigma$ and $2\sigma$ confidence regions, respectively.
    The $1\sigma$ uncertainties are: $\sigma_{m}/m=2.2\times 10^{-4}$, $\sigma_{ga}/ga=0.27$, $\sigma_{\theta}/\theta=0.79$, and $\sigma_{\phi}/\phi=1.13$.}
    \label{fig:Fisher}
\end{figure*}

Neglecting frequency evolution, a monochromatic GW from a binary inspiral is characterized by seven parameters: the frequency $f$, overall amplitude $A$, source location $\hat{k}$, 
inclination $\iota$, polarization angle $\psi$, and orbital phase $\varphi_0$ at a given time.
Here, for illustration we fix the inclination, polarization angle, and orbital phase, and explore the uncertainties in the remaining four parameters: the binary's orbital radius $R$ and luminosity distance $d_L$, which are equivalent to the frequency $f$ and amplitude $A$ given the component masses of binary, and the angels $\theta$ and $\phi$ specifying the source location. 
We numerically evaluate the matrix elements and invert the matrix.
In Fig.~\ref{fig:Fisher_GW}, we show the posterior distributions for a binary system with $\mathrm{SNR}=\sqrt{(h\mid h)}\approx 8.68$, assuming a four-year observation and the detector parameters specified in Appendix~\ref{ap:detector}.

% Parameter correspondence
A vector ULDM is also characterized by seven parameters: the ULDM mass $m$ and three complex amplitudes $a_p$; see Eq.~(\ref{eq:DM mono}). 
However, the amplitudes can be recast into a set of quantities with clearer geometric meaning. 
For times shorter than the coherence time, the endpoint of $\mathbf{A}$ at a fixed spatial point $\mathbf{x}_0$ traces out an ellipse (Appendix~\ref{ap:ULDM}). The ellipse is characterized by the normal vector of its plane $\hat{n}$, the semi-major axis $a = \max\left[\left|\mathbf{A}(t,\mathbf{x}_0)\right|\right]$, eccentricity $e$, and an angle $\psi$ specifying the rotation of the ellipse around $\hat{n}$. The position of the endpoint within the ellipse at $t=0$ is specified by an initial phase $\varphi_0$.
Comparing with the GW parameters, a mapping emerges immediately:
\begin{equation} \label{eq:para mapping}
    f \rightarrow m,\,
    A \rightarrow a,\,
    \hat{k} \rightarrow \hat{n},\,
    (\iota,\psi) \rightarrow (e,\psi),\,
    \varphi_0 \rightarrow \varphi_0.
\end{equation}
This implies that the prior volumes of GW and ULDM signals are nearly the same. 
Consequently, the preference between signal models in data analysis is largely driven by their respective likelihoods.
% Analogous to the GW case
As a first step toward ULDM parameter estimation, we assume that the eccentricity $e$, the angle $\psi$, and the initial phase $\varphi_0$ are known, and investigate the inference uncertainties in the remaining parameters: the ULDM mass $m$, the dimensionless effective signal strength $ga$, and the polar angle $\theta$ and azimuth angle $\phi$ specifying the normal vector of the elliptical plane. 
The results are shown in Fig.~\ref{fig:Fisher_DM}, where the ULDM signal has a SNR of $6.21$.
As shown, the ULDM mass can be measured with high precision, achieving a relative uncertainty of $10^{-4}$.

Comparing the results, we find that the direction of the normal vector of the elliptical plane for ULDM is not as well constrained as the source location in the GW case.
This can be understood as follows:
the angular resolution is roughly given by $\lambda / r_{\odot}$ for GWs while $\lambda_c / r_{\odot}$ for ULDM. 
At the same frequency, the ULDM coherence length $\lambda_c$ is larger than the GW wavelength $\lambda$ by a factor of approximately $\sigma^{-1} \sim 10^3$.
Thus, the angular resolution for ULDM is much worser than that for GWs.

%------------------------------------------------------------
\section{Conclusion}\label{sec:conl}
We have identified a characteristic spectral structure to distinguish quasi-monochromatic GW signals from those induced by ULDM. 
%based on their modulated spectra arising from the detector's heliocentric motion.
% Explain modulation
The motion of the detector modulates the signal, splitting the monochromatic signal into a set of harmonics separated by the orbital frequency.
% What we do
The distinct couplings to the detector and kinetic properties of GWs and ULDM would lead to distinct harmonic structures, which provides a way for discrimination.
% Fisher 
Using the Fisher matrix formalism, we have quantified how well signal parameters can be extracted from observations.
Our results show that the parameters of ULDM, such as its mass, can be determined with high precision.

To capture the key features, we have illustrated the modulation effects using an analytical Keplerian orbit.
While the main spectral characteristics are likely robust, the more sophisticated numerical orbits may give rise to further intricate spectral structures, which can be explored in future.
Meanwhile we have shown that the spectra exhibit qualitative differences and discussed semi-quantitatively their implications for detection, a full Bayesian analysis is still needed to address the model selection problem in the presence of detector noise. Moreover, given that the typical SNR for sources such as galactic white dwarf binaries is only a few,
exploring the full posterior distributions of all parameters also requires a Monte Carlo approach.

%------------------------------------------------------------
\section*{acknowledgement}
This work is supported by the National Key Research and Development Program of China (Grant No.~2021YFC2201901), and the Fundamental Research Funds for the Central Universities. 

Note added: While we were submitting the manuscript, a preprint~\cite{Gue:2025iab} discussing scalar dark matter appeared on arXiv.

%------------------------------------------------------------
\appendix
%------------------------------------------------------------
\section{Keplerian orbits} \label{ap:orbit}
We adopt analytic Keplerian orbits for spacecrafts~\cite{PhysRevD.69.082003}:
\begin{align} \label{eq:Keplerian orbit}
   & x_i(t) = r_{\odot}\cos\alpha + \frac{1}{2}er_{\odot}\left[\cos(2\alpha-\beta_i)-3\cos\beta_i\right] \nonumber\\
    &
    + \frac{1}{8} e^2 r_{\odot}[3 \cos (3 \alpha-2 \beta_i)-10 \cos\alpha-5 \cos (\alpha-2 \beta_i)], \nonumber\\
   & y_i(t) = r_{\odot}\sin\alpha + \frac{1}{2}er_{\odot}\left[\sin(2\alpha-\beta_i)-3\sin\beta_i\right] \nonumber\\
    &
    + \frac{1}{8} e^2 r_{\odot}[3 \sin (3 \alpha-2 \beta_i)-10 \sin\alpha+5 \sin (\alpha-2 \beta_i)], \nonumber\\
  &  z_i(t) = -\sqrt{3}er_{\odot}\cos(\alpha-\beta_i) \nonumber\\
    &+ \sqrt{3} e^2 r_{\odot}\left[\cos ^2(\alpha-\beta_i)+2 \sin ^2(\alpha-\beta_i)\right],
\end{align}
where $\mathbf{x}_i = [x_i, y_i, z_i]$ ($i=1,2,3$),
$r_{\odot} = 1\text{AU}$ is the distance between the constellation's guiding center and the Sun,
$e$ is the orbital eccentricity, $\alpha = 2\pi f_m t + \kappa$ is the orbital phase of the constellation's guiding center with $f_m = 1/\text{yr}$, $\beta_i=2\pi(i-1)/3+\lambda$ are the relative phases of spacecraft within the constellation,
and the parameters $\kappa$ and $\lambda$ specify the positions of spacecraft at $t=0$.
The coordinate system is chosen such that the Sun is at the origin, the $x$-axis points toward the vernal equinox, the $z$-axis is perpendicular to the ecliptic, and the $y$-axis lies in the ecliptic plane, forming a right-handed coordinate system.
From Eq.~(\ref{eq:Keplerian orbit}), the distances between spacecraft are $|\mathbf{x}_i(t) - \mathbf{x}_j(t)| = 2\sqrt{3}er_{\odot} + \mathcal{O}(e^2)$.
Since the interferometer's arm length is around a few million kilometers, we have $e \sim 5\times 10^{-3}$ and the arm lengths fluctuate on $0.1\%$ level over the course of a year.

The exceptional stability of the arm lengths allows the use of the rigid adiabatic approximation~\cite{PhysRevD.69.082003}, which, as shown later, greatly simplifies the derivation of detector response and works very well.
The approximation consists of two components: ``adiabatic" and ``rigid".
The first part ``adiabatic" comes from the fact that over timescales of a few hundred seconds, the spacecraft displacements are sufficiently small that the constellation can be approximated as quasi-static. Accordingly, the detector response within each interval can be evaluated using fixed spacecraft positions. 
The second component, ``rigid", relies on the stability of the interferometer’s arm lengths. It therefore assumes that all arm lengths remain constant and equal to a fiducial arm length.

%------------------------------------------------------------
\section{GW signal} \label{ap:GW}
\subsection{Monochromatic GWs}
A GW tensor can be decomposed into the two polarization modes~\cite{Maggiore:2007ulw}:
\begin{equation} \label{eq:GW tensor}
    \boldsymbol{h}(t) = h_p(t)\mathbf{e}^p(\hat{k}) ,
\end{equation}
where the polarization tensors in the solar system barycenter~(SSB) frame are defined as 
\begin{equation}
    \mathbf{e}^+ = \hat{u}\otimes\hat{u} - \hat{v}\otimes\hat{v},\quad
    \mathbf{e}^{\times} = \hat{u}\otimes\hat{v} + \hat{v}\otimes\hat{u},
\end{equation}
where, $\hat{k} = -[\cos\phi\sin\theta,\sin\phi\sin\theta,\cos\theta]$ is the unit wave vector for a source located at polar angle $\theta$ and azimuth angle $\phi$, $\hat{u} = -\partial_{\theta}\hat{k}$, and $\hat{v} = \partial_{\phi}\hat{k}/\sin\theta$.
$h_p(t)$ denotes the time-domain GW waveforms, which is related to that in the radiation frame by $h_{\times} \pm ih_{+} = e^{\mp 2i\psi}(h^{\text{rad}}_{\times} \pm ih^{\text{rad}}_{+})$ with $\psi$ the polarization angle.

For monochromatic GWs from inspiral binaries, 
the waveforms are given by
\begin{align}
    h^{\textrm{rad}}_{+}(t) &= A (1 + \cos^2\iota)\cos(2\pi ft + \varphi_0),\nonumber \\
    h^{\textrm{rad}}_{\times}(t) &= 2A \cos\iota\sin(2\pi ft + \varphi_0) ,
\end{align}
where $f$ is the GW frequency, $A$ the overall amplitude, $\iota$ the inclination, 
and $\varphi_0$ the orbital phase.
We express the waveforms in the SSB frame as
\begin{equation} \label{eq:GW mono waveform}
    h_p(t) = H_p e^{i2\pi ft}, 
\end{equation}
where the complex amplitudes $H_p$ are functions of $A$, $\iota$, $\psi$, and $\varphi_0$.

\subsection{Response to GWs}
We now derive the signal induced by a monochromatic GW in interferometers.
Since the data channels (e.g. Eq.~(\ref{eq:def TDI X})) are constructed from the single-link data streams, we first investigate the response of a single laser link to the GW.
Substituting Eqs.~(\ref{eq:GW tensor}) and (\ref{eq:GW mono waveform}) into Eq.~(\ref{eq:1link GW signal}), we have 
\begin{equation} \label{eq:1link GW signal decompose}
    y_{rs}(t) 
    = e^{i2\pi ft}\sum_{p=+,\times} R^p_{rs}(f,\hat{k})H_p , 
\end{equation}
where we have defined the GW antenna patterns~\cite{Romano:2016dpx} of a laser link as
\begin{equation} \label{eq:1link GW ant}
    R^p_{rs}(f,\hat{k}) = G^p_{rs}(\hat{k}) \mathcal{T}_{rs}(f,\hat{k})
     e^{-i2\pi f \hat{k}\cdot\mathbf{x}_r} .
\end{equation}
Here $G^p_{rs}$ are the frequency-independent geometric factors
\begin{equation} \label{eq:1link GW geo factor}
    G^p_{rs}(\hat{k}) = 
    \frac{1}{2}\hat{n}_{rs}:\mathbf{e}^p(\hat{k}):\hat{n}_{rs}.
\end{equation}
And $\mathcal{T}_{rs}$ denotes the polarization-independent transfer function
\begin{equation} \label{eq:1link GW transfer func}
\begin{split}
    \mathcal{T}_{rs}(f,\hat{k}) 
    =& -(i2\pi fL_{rs})\text{sinc}\left(\pi fL_{rs}(1-\hat{k}\cdot\hat{n}_{rs})\right) \\
    & \times e^{-i\pi fL_{rs}(1-\hat{k}\cdot\hat{n}_{rs})} ,
\end{split}
\end{equation}
where $L_{rs}=|\mathbf{x}_r-\mathbf{x}_s|$, and $\text{sinc}(x) = \sin x/x$.
The transfer function essentially measures the phase difference of GW between the spacetime points where laser is emitted and received.

Next we proceed to derive the $X$-channel signal. Using Eqs.~(\ref{eq:1link GW signal decompose}) and (\ref{eq:def TDI X}), we have
\begin{equation} \label{eq:X GW signal}
    X(t) = e^{i2\pi ft}\sum_{p=+,\times} R^p(f,\hat{k}) H_p  ,
\end{equation}
where $R^p(f,\hat{k})$ are the GW antenna patterns of $X$ channel.
We adopt the rigid adiabatic approximation introduced in Appendix~\ref{ap:orbit}, under which $L_{rs} = L$ for $\forall r,s$, where $L$ is the detector's fiducial arm length. Consequently, $R^p(f,\hat{k})$ only depends on frequency through the combination $\delta = 2\pi fL$.
As we focus on signals with frequencies below $10~\text{mHz}$, the detector operates in the small-antenna limit $\delta \ll 1$~\cite{Romano:2016dpx}.
This allows for a series expansion of $R^p(f,\hat{k})$ in $\delta$.
Keeping the leading order, we have
\begin{equation} \label{eq:X GW ant low}
    R^p(f,\hat{k}) \simeq 4\delta^2 G^p(\hat{k}) e^{-i2\pi f\hat{k}\cdot\mathbf{x}_1},
\end{equation}
where $G^p(\hat{k}) = G^p_{12}(\hat{k}) - G^p_{13}(\hat{k})$, depending on time through the unit arm direction $\hat{n}_{12}$ and $\hat{n}_{13}$.
For the analytical Keplerian orbits given in Appendix~\ref{ap:orbit}, we find that
\begin{equation} \label{eq:Mich GW geo factor exp}
    G^{p}(t;\hat{k}) = \sum_{n=-4}^{4} G^{(n)}_p(\hat{k}) e^{i2\pi nf_m t} ,
\end{equation}
where the geometric coefficients $G^{(n)}_p$ are listed in Table~\ref{tab:GW harm cof}.
Since $G^{p}(t;\hat{k})$ is real, we have $G^{(n)}_p(\hat{k}) = \left(G^{(-n)}_p(\hat{k})\right)^*$.
The phase term is approximately given by $-2\pi f \hat{k}\cdot \mathbf{x}_1\simeq2\pi f r_{\odot}\sin\theta \cos(\alpha-\phi)$, where $\alpha = 2\pi f_m t + \kappa$ is the orbital phase of the constellation's guiding center.
Using the Jacobi-Anger identity, we can expand the phase into harmonics of $f_m$:
\begin{equation} \label{eq:Doppler exp}
    e^{-i2\pi f \hat{k}\cdot \mathbf{x}_1(t)} = \sum_{n=-\infty}^{n=\infty} \mathcal{D}^{(n)}(f,\hat{k}) e^{i2\pi nf_mt} ,
\end{equation}
where the Doppler coefficients are given by
\begin{equation} \label{eq:doppler coef}
    \mathcal{D}^{(n)}(f,\hat{k}) = J_n(\epsilon) e^{in(\kappa-\phi+\pi/2)} ,
\end{equation}
where $\epsilon = 2\pi fr_{\odot}\sin\theta$, and $J_n$ is the $n$-th Bessel function of the first kind.
Substituting Eqs.~(\ref{eq:Mich GW geo factor exp}) and (\ref{eq:Doppler exp}) into Eq.~(\ref{eq:X GW signal}), we arrive at Eq.~(\ref{eq:X GW modul}) used in the main text.

Although in general asymmetric about the zeroth harmonic, the spectra in some cases are symmetric when the following condition is satisfied, using Eqs.~(\ref{eq:X GW modul}) and (\ref{eq:doppler coef})
\begin{equation}
\begin{split}
    & \sum_{m,n} J_{l-m}J_{l-n} e^{-i(m-n)(\kappa-\phi+\pi/2)} \sum_{p,q} H_pH^*_q \\
    & \times\left(G^{(m)}_{p}G^{(-n)}_{q} - (-1)^{m+n}G^{(-n)}_{p}G^{(m)}_{q}\right) = 0,\; 
    \forall l.
\end{split}
\end{equation}
When Doppler modulation is insignificant, i.e., $\epsilon < 1$, it reduces to
\begin{equation}
   \textrm{Im}(H_+H^*_{\times}) \left(G^{(n)}_{+}G^{(-n)}_{\times} - G^{(-n)}_{+}G^{(n)}_{\times}\right) = 0,\; \forall n.
\end{equation}
%which is not generally met for sources at arbitrary sky locations.

\begin{table*}[htbp]
\centering
\caption{Geometric coefficients for GWs with the analytical Keplerian orbits.
The orbital parameters $\kappa$ and $\lambda$ are defined in Appendix~\ref{ap:orbit}.
The coefficients with negative indices are given by $G^{(n)}_{p}(\hat{k}) = \left(G^{(-n)}_{p}(\hat{k})\right)^*$.}
\label{tab:GW harm cof}
\setlength{\tabcolsep}{10pt}
\renewcommand{\arraystretch}{1.5}
\begin{tabular}{c|ccc}
\toprule
$n$ &
$G^{(n)}_+$ &
$G^{(n)}_\times$ &\\
\midrule
0 &
$-\frac{9\sqrt{3}}{128} [3 + \cos(2\theta)]\sin(2(\phi - \lambda))$ &
$\frac{9\sqrt{3}}{32} \cos\theta \cos(2(\phi - \lambda))$ \\
1 &
$-\frac{9i}{64} \sin(2\theta) e^{i(\kappa + \phi - 2\lambda)}$ &
$-\frac{9}{32} \sin\theta e^{i(\kappa + \phi - 2\lambda)}$ \\
2 &
$\frac{9\sqrt{3}i}{128} [1-\cos(2\theta)] e^{i2(\kappa-\lambda)}$ &
0  \\
3 &
$\frac{3i}{64} \sin(2\theta) e^{i(3\kappa-2\lambda-\phi)}$ &

$-\frac{3}{32} \sin\theta e^{i(3\kappa-2\lambda-\phi)}$ \\
4 &
$\frac{\sqrt{3}i}{256} [3+\cos(2\theta)] e^{2i(2\kappa-\lambda-\phi)}$ &
$-\frac{\sqrt{3}}{64} \cos\theta e^{2i(2\kappa-\lambda-\phi)}$ \\
\bottomrule
\end{tabular}
\end{table*}

%------------------------------------------------------------
\section{ULDM signal} \label{ap:ULDM}
\subsection{Stochastic ULDM}
The ULDM field is modeled as a superposition of plane waves,
each representing the collective behavior of ULDM particles residing in the same volume of phase space~\cite{PhysRevA.97.042506, PhysRevD.97.123006, PhysRevD.111.015028, Kim:2023pkx, PhysRevD.110.095015}:
\begin{equation} \label{eq:DM rand field}
    \mathbf{A}(x) 
    = A_0  
    \sum_p \hat{x}^p \sum_{\mathbf{v}} \beta_{\mathbf{v},p}\sqrt{\Delta(\mathbf{v})} 
    e^{i\left(\omega t - \mathbf{k} \cdot \mathbf{x} + \varphi_{\mathbf{v},p}\right)} ,
\end{equation}
where $\omega\simeq m(1 + |\mathbf{v}|^2/2)$, $\mathbf{k} \simeq m\mathbf{v}$,
$A_0$ is a normalization constant fixed by the local DM energy density, $\Delta(\mathbf{v}) = d\mathbf{v}f(\mathbf{v})$, $f(\mathbf{v})$ is the normalized velocity distribution of DM.
Each plane wave has a random amplitude $\beta_{\mathbf{v},p}$ drawn independently from the Rayleigh distribution with unit scale parameter, and a random phase $\varphi_{\mathbf{v},p}$ drawn uniformly over $[0, 2\pi)$.

By factoring out the rapid Compton oscillation term, we recover Eq.~(\ref{eq:DM mono}), where the complex amplitudes $a_p(x)$ denote the sum over the velocity-dependent terms and vary over the coherence time $\tau_c = 2\pi/m\sigma^2 \sim 10^6f^{-1}_c$, given a typical DM velocity dispersion $\sigma \sim 10^{-3}$.
For shorter times, the amplitude's evolution can be ignored, and the field at a fixed spatial point is described by 
\begin{equation} \label{eq:DM mono co}
    \mathbf{A}(t) = \boldsymbol{a}e^{imt} ,
\end{equation}
where $\boldsymbol{a}$ is a constant complex vector.
The square of $\boldsymbol{a}$ is a complex number and can be written as $\boldsymbol{a}^2 = \rho e^{2i\alpha}$ with $\rho$ real.
If we define another complex vector $\bar{\boldsymbol{a}} = \boldsymbol{a}e^{-i\alpha}$, then $\bar{\boldsymbol{a}}^2= \rho$.
We can decompose $\bar{\boldsymbol{a}}$ into two real vectors $\bar{\boldsymbol{a}} = \bar{\boldsymbol{a}}_1 - i \bar{\boldsymbol{a}}_2$.
Since $\bar{\boldsymbol{a}}^2 = \bar{\boldsymbol{a}}^2_1 - \bar{\boldsymbol{a}}^2_2 - 2i\bar{\boldsymbol{a}}_1\cdot\bar{\boldsymbol{a}}_2$ is real, we have $\bar{\boldsymbol{a}}_1\cdot\bar{\boldsymbol{a}}_2 = 0$.
Therefore, the field can be expressed as
\begin{equation}
    \mathbf{A}(t) = \bar{\boldsymbol{a}}_1 \cos(mt - \alpha)
    + \bar{\boldsymbol{a}}_2 \sin(mt - \alpha),
\end{equation}
where $\bar{\boldsymbol{a}}_1$ and $\bar{\boldsymbol{a}}_2$ are two orthogonal vectors. The endpoint of $\mathbf{A}(t)$ traces out an ellipse.

\subsection{Response to ULDM}
Since the field is a superposition of plane waves, 
the overall signal is the sum of contributions from each plane wave.
Using Eq.~(\ref{eq:1link DM signal}), we find the contribution from a plane wave is given by
\begin{equation} \label{eq:1link DM signal decompose}
    y_{rs}(t) = e^{i2\pi ft}\sum_{p} \mathcal{R}^p_{rs}(f,\mathbf{v})A_p ,
\end{equation}
where $f = \omega/2\pi$ and $A_p = A_0\beta_{\mathbf{v},p}\sqrt{\Delta(\mathbf{v})}e^{i\varphi_{\mathbf{v},p}}$.
We define the ULDM antenna patterns of a laser link as
\begin{equation} \label{eq:1link DM ant}
    \mathcal{R}^p_{rs}(f,\mathbf{v}) = \mathcal{G}^p_{rs}
    \mathfrak{T}_{rs}(f,\mathbf{v}) 
    e^{-i\mathbf{k}\cdot\mathbf{x}_r} ,
\end{equation}
where the geometry factors are given by 
\begin{equation} \label{eq:1link DM geo factor}
    \mathcal{G}^p_{rs} = \hat{n}_{rs}\cdot\hat{x}^p ,
\end{equation}
and the polarization-independent transfer function
\begin{equation} \label{eq:1link DM transfer func}
    \mathfrak{T}_{rs}(f,\mathbf{v}) = 
    2i\sin\left(\pi fL_{rs}(1-\mathbf{v}\cdot\hat{n}_{rs})\right)
    e^{-i\pi fL_{rs}\delta(1-\mathbf{v}\cdot\hat{n}_{rs})}.
\end{equation}
%which depends the ULDM phases at the spacetime points where laser is emitted and received.

Using Eqs.~(\ref{eq:DM rand field}), (\ref{eq:1link DM ant}) and (\ref{eq:def TDI X}), the overall signal in $X$ channel is given by
\begin{equation} \label{eq:Mich DM signal}
    X(t) = \sum_{p,\mathbf{v}} A_p\mathcal{R}^p(f, \mathbf{v}) e^{i\omega t},
\end{equation}
where $\mathcal{R}^p$ are the ULDM antenna patterns of $X$ channel, characterizing its response to a ULDM plane wave with frequency $f$ and velocity $\mathbf{v}$.
In the low-frequency limit, we have
\begin{equation} \label{eq:Mich DM ant}
\begin{split}
    \mathcal{R}^p(f, \mathbf{v}) 
    \simeq& -2\delta^2 \left[ i\delta \mathcal{G}^p \right.\\
    &\left. - 2 \big[(\mathbf{v}\cdot\hat{n}_{12})\mathcal{G}^p_{12} - (\mathbf{v}\cdot\hat{n}_{13})\mathcal{G}^p_{13}\big]
    \right] e^{-i\mathbf{k} \cdot \mathbf{x}_1} ,
\end{split}
\end{equation}
where $\delta = 2\pi fL$, and $\mathcal{G}^p = \mathcal{G}^p_{12} - \mathcal{G}^p_{13}$. 
For the analytical Keplerian orbits given in Appendix~\ref{ap:orbit}, we find
\begin{equation} \label{eq:Mich DM geo factor exp}
    \mathcal{G}^p(t) = \sum_{n=-2}^{2} \mathcal{G}_p^{(n)} e^{i2\pi nf_m t} ,
\end{equation}
where the coefficients $\mathcal{G}_p^{(n)}$ are listed in Table~\ref{tab:DM harm cof}.
The velocity-dependent terms in Eq.~(\ref{eq:Mich DM ant}) represent the contribution of the field gradient and only become important when $\delta \leq |\mathbf{v}| \sim \sigma \sim 10^{-3}$. 
Since LISA/Taiji's low-frequency cutoff is around $0.1~\textrm{mHz}$, we have $\delta > |\mathbf{v}|$, and the velocity-dependent terms can be ignored.
Thus, Eq.~(\ref{eq:Mich DM signal}) reduces to
\begin{equation}
    X(t) \simeq   -2i(mL)^3e^{imt}\sum_p \mathcal{G}^p(t) a_p(t,\mathbf{x}_1) ,
\end{equation}
where we approximate $\delta = mL(1+|\mathbf{v}|^2/2) \approx mL$ and use the field description given in Eq.~(\ref{eq:DM mono}).
Substituting Eq.~(\ref{eq:Mich DM geo factor exp}), we recover Eq.~(\ref{eq:X DM modul}).

\begin{table*}[htbp]
\centering
\caption{Geometric coefficients for ULDM with the analytical Keplerian orbits.
The polarization modes are defined with respect to the three unit basis vectors: $\hat{x}^p = \hat{x},\hat{y},\hat{z}$.
The coefficients with negative indices are given by $\mathcal{G}^{(n)}_{p}(\hat{k}) = \left(\mathcal{G}^{(-n)}_{p}(\hat{k})\right)^*$.}
\label{tab:DM harm cof}
\setlength{\tabcolsep}{10pt}
\renewcommand{\arraystretch}{1.5}
\begin{tabular}{c|cccc}
\toprule
$n$ &
$\mathcal{G}^{(n)}_x$ &
$\mathcal{G}^{(n)}_y$ &
$\mathcal{G}^{(n)}_z$ \\
\midrule
0 &
$-\frac{3}{4}\sin\lambda$ &
$\frac{3}{4}\cos\lambda$ &
0 \\
1 &
0 &
0 &
$-\frac{\sqrt{3}i}{4}e^{i(\kappa-\lambda)}$ \\
2 &
$\frac{i}{8}e^{i(2\kappa-\lambda)}$ &
$\frac{1}{8}e^{i(2\kappa-\lambda)}$ &
0 \\
\bottomrule
\end{tabular}
\end{table*}

%------------------------------------------------------------
\section{Finite signal duration} \label{ap:finite time}
Due to the limited observation time, the recorded signal has a finite duration and is given by
\begin{equation} \label{eq:finite signal}
    s_w(t) = s(t)w(t),
\end{equation}
where $s(t)$ is the ideal signal with infinite duration, and $w(t)$ is the rectangular window function:
\begin{equation}
    w(t) = \Theta(t)\Theta(T-t),
\end{equation}
where $\Theta(x)$ is the Heaviside step function, and we assume that the observation is performed between $t=0$ and $t=T$.
Fourier transforming Eq.~(\ref{eq:finite signal}), we have
\begin{equation} \label{eq:window Fourier coef}
    \tilde{s}_w(f) = \int^{\infty}_{-\infty} df' \tilde{s}(f') \tilde{w}(f-f'),
\end{equation}
where the window function's Fourier transform is given by
\begin{equation}
    \tilde{w}(f) = T e^{-i\pi fT} \text{sinc}(\pi fT).
\end{equation}
For the signals given by Eqs.~(\ref{eq:X GW modul}) and (\ref{eq:X DM modul}), we have
\begin{equation} \label{eq:para signal spectrum}
    \tilde{s}(f) = \sum_n C_n \delta\left(f-f_n\right),
\end{equation}
where $\delta(x)$ denotes the Dirac delta function, and $f_n = f_0 + nf_m$ with $f_0$ the signal frequency without modulation. 
Substituting Eq.~(\ref{eq:para signal spectrum}) into Eq.~(\ref{eq:window Fourier coef}), we have
\begin{equation} \label{eq:window signal spectrum}
    \tilde{s}_w(f) = T\sum_n C_ne^{-i\pi (f-f_n)T} \text{sinc}\left(\pi (f-f_n)T\right),
\end{equation}
which is shown in Fig.~\ref{fig:dm2gw_X} and agrees well with numerical spectra obtained in  Appendix~\ref{ap:simulation}.

%------------------------------------------------------------
\section{Numerical simulation}~\label{ap:simulation}
Now we discuss numerical simulation of ULDM configuration for the interferometer. 
We first simulate the random ULDM field in time domain for a given DM velocity dispersion, compute its PSD and compare with the theoretical prediction.
We then present the simulation of the signal in a moving detector without the approximation adopted in the analytical derivation.
We compare the analytical signal spectra with those obtained from numerical simulations and find good agreement.

\subsection{ULDM field}
The basic expression used in our ULDM field simulation is Eq.~(\ref{eq:DM rand field}):
\begin{equation*}
    \mathbf{A}(x) 
    = A_0  
    \sum_p \hat{x}^p \sum_{\mathbf{v}} \beta_{\mathbf{v},p}\sqrt{\Delta(\mathbf{v})} 
    e^{i\left(\omega t - \mathbf{k} \cdot \mathbf{x} + \varphi_{\mathbf{v},p}\right)}.
\end{equation*}
We outline two general considerations underlying the simulation.
First, while the velocity distribution of DM is continuous, only a finite number of points can be sampled in momentum space on a computer. Therefore, we carry out the simulation in a cubic box of side length $L$ and impose the periodic boundary condition on the field, i.e. $A_i(t,\mathbf{x}+L\hat{x}^j) = A_i(t,\mathbf{x})$, where $\hat{x}^j$ ($j=1,2,3$) denote the three unit vectors along the edges of the box.  
This results in quantized momenta $k_i = 2\pi n_i/L$ with $n_i=0,\pm1,\pm2,\cdots$, and the angular frequency follows from the nonrelativistic dispersion relation $\omega \simeq m + |\mathbf{k}|^2/2m$.
Second, since the characteristic time and length scales of the field are the Compton time $f_c^{-1}$ and the coherence length $\lambda_c$, it is convenient to normalize time and length to these scales.
We define the following dimensionless quantities in our simulation: $\bar{t} = f_ct$, $\bar{\mathbf{x}} = \mathbf{x}/\lambda_c$, $\bar{L} = L/\lambda_c$, and $\bar{k}_i = \lambda_ck_i/2\pi = n_i/\bar{L}$.

For illustration, here we adopt an isotropic Maxwell distribution for DM
\begin{equation} \label{eq:iso Maxwell}
        f(\mathbf{v})=\frac{1}{(2\pi\sigma^2)^{3/2}} e^{-\frac{\mathbf{v}^2}{2\sigma^2}} \Longrightarrow
        f(\bar{\mathbf{k}})=\frac{1}{(2\pi)^{3/2}} e^{-\frac{\bar{\mathbf{k}}^2}{2}},
    \end{equation}
where $\sigma \sim 10^{-3}$.
The finite box size corresponds to a finite resolution $\Delta\bar{k}_i = \bar{L}^{-1}$ in momentum space. To ensure a smooth sampling of the momentum distribution, the resolution needs to satisfy $\left|\nabla_{\bar{\mathbf{k}}}f\right|  \ll \bar{L}$. For the isotropic Maxwell distribution given in Eq.~(\ref{eq:iso Maxwell}), we find that a box size $\bar{L} \geq 6$ is sufficient.
Additionally, we also require the momentum-space box to be sufficiently large so that the distribution can be properly covered. Define the covering ratio as $\eta = \bar{L}^{-3}\sum_{\mathbf{k}} f(\bar{\mathbf{k}})$, $\eta \geq 99\%$ is achieved by $\bar{k}_{i,\max} \geq 3$. 

The values of $\bar{L}$ and $\bar{k}_{i,\max}$ define a grid in momentum space.
At each grid point, we generate a plane wave and attach to it a random amplitude and phase according to Eq.~(\ref{eq:DM rand field}).  
Since the field's polarizations are uncorrelated, the amplitude and phase are sampled independently for each polarization. 
We simulate the field on a spacetime grid, with the time and spatial steps determined by the sampling theorem:
to resolve the field's temporal evolution, the temporal sampling frequency must satisfy $f_s^t \geq 2 f_{\max} = 2f_c(1 + |\bar{\mathbf{k}}_{\max}|^2\sigma^2/2) \simeq 2f_c$; while a spatial sampling frequency $f_s^x \geq  k_{i,\max}/\pi$ is required to ensure a adequate spatial resolution.

\begin{figure}
    \includegraphics[scale=1.1]{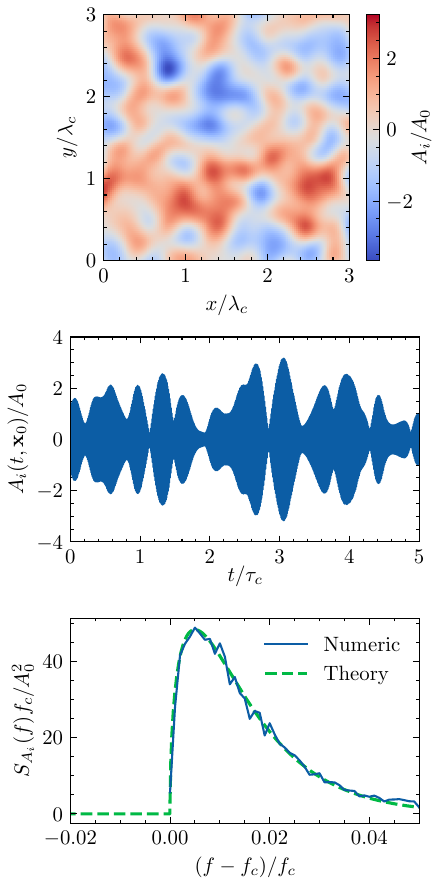}
    \caption{Simulation of the random ULDM field. Top: a 2D snapshot of one polarization mode of the field. Middle: temporal evolution of the same polarization mode at a fixed spatial point. 
    Bottom: the field's PSD (solid blue line) computed from the time-domain signal and the theoretical prediction (dashed green line) given by Eq.~(\ref{eq:ULDM psd}).
    For illustration, a high velocity dispersion $\sigma=0.1$ is used.
    The PSD is averaged over $10^3$ field realizations.
    }
    \label{fig:uldm simul}
\end{figure}

In the top and middle panels of Fig.~\ref{fig:uldm simul}, we show the evolution of one polarization mode of the simulated field, where its amplitude fluctuates over the coherence time and coherence length.
The bottom panel presents the power spectral density~(PSD) computed from the time-domain signal, in comparison with the theoretical prediction~\cite{PhysRevD.110.095015} 
\begin{equation} \label{eq:ULDM psd}
    S_{A_i}(f) 
        = \frac{4\pi^2 A^2_0}{m} v f(v),
\end{equation}
where $v = \sqrt{2(f/f_c-1)}$.

\subsection{Moving detector}
In modeling the detector motion, the key step is to solve the transcendental light propagation equation $t-t_s = |\mathbf{x}_r(t)-\mathbf{x}_s(t_s)|$, which determines the light travel time $L_{rs} = t-t_s$ between spacecraft.
The equation can be solved perturbatively in powers of $\varepsilon = R_S/2r_{\odot} \sim 10^{-8}$~\cite{PhysRevD.72.122003}, where $R_S = 2GM_{\odot}$ is the Schwarzschild radius of the Sun,
\begin{equation}
    L_{rs} \simeq L_{rs}^{(0)} + L_{rs}^{(1/2)} + \mathcal{O}(\varepsilon^1) .
\end{equation}
The zeroth-order term is the distance between the two spacecraft at $t$
\begin{equation}
    L_{rs}^{(0)}(t) = |\mathbf{x}_r(t)-\mathbf{x}_s(t)| . 
\end{equation}
The half-order term accounts for the spacecraft motion and is given by
\begin{equation}
    L_{rs}^{(1/2)}(t) = \mathbf{v}_r(t)\cdot\left(\mathbf{x}_r(t)-\mathbf{x}_s(t)\right) ,
\end{equation}
where $\mathbf{v}_r(t)$ is the velocity of the receiving spacecraft at $t$. 
The higher order terms, such as $L_{rs}^{(1)}$, include additional corrections from the spacecraft motion and general relativistic effects on light propagating in the Sun's gravitational field. They contribute a correction of $L_{rs}^{(1)} \sim 10^{-7}~\textrm{s}$, which is about four orders of magnitude smaller than $L_{rs}^{(1/2)}$.
In our simulation, we take the light travel time as $ L_{rs} = L_{rs}^{(0)} + L_{rs}^{(1/2)}$.
The unit vector $\hat{n}_{rs}$ can also be obtained perturbatively, and we adopt $\hat{n}_{rs} = \hat{n}_{rs}^{(0)} + \hat{n}_{rs}^{(1/2)}$, which incorporates the point-ahead effect due to the spacecraft motion.
Moreover, for time-dependent arm lengths, the time-delayed operation in Eq.~(\ref{eq:def TDI X}) should take a nested form $ y_{rs,jk}(t) = y_{rs}(t - L_k(t) - L_j(t-L_k(t)))$, which is implemented in the simulation.

\begin{figure*}
    \begin{center}
        \includegraphics[scale=0.55]{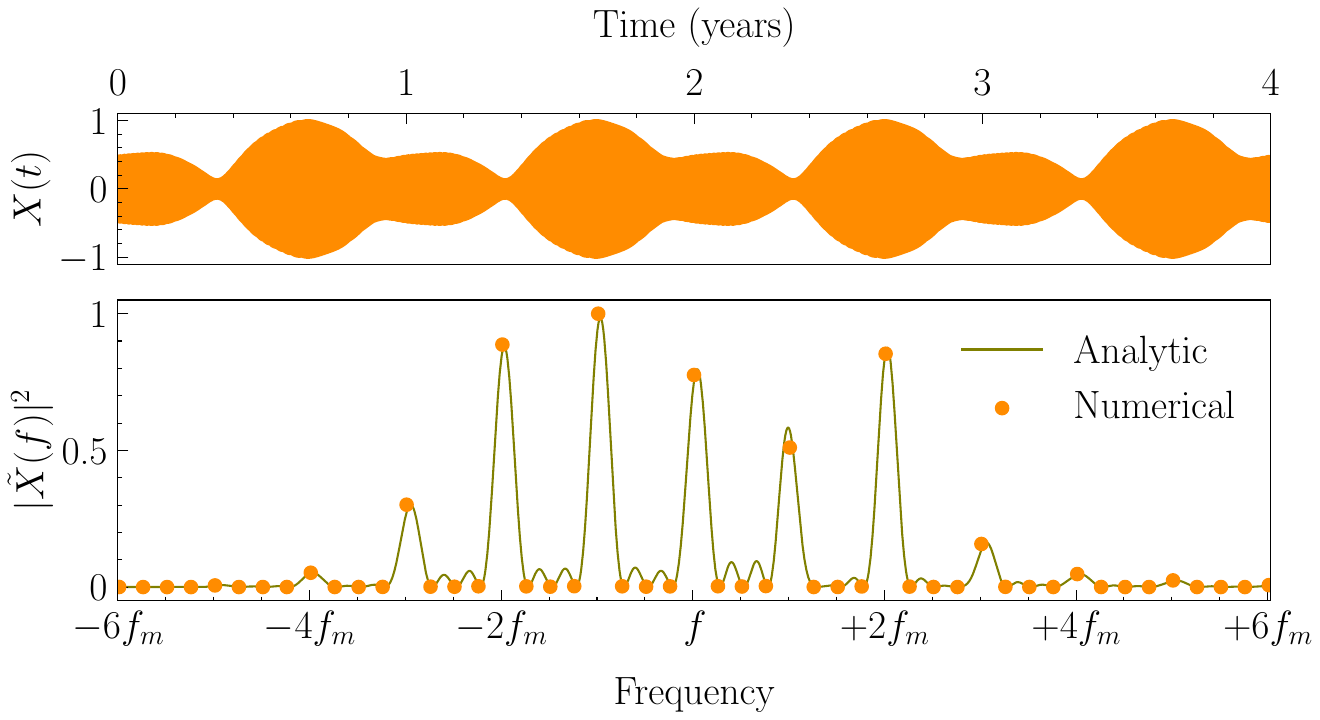}\label{fig:gwtf}
        %\\ (a) Waveform and spectrum of GW.\\
        \includegraphics[scale=0.55]{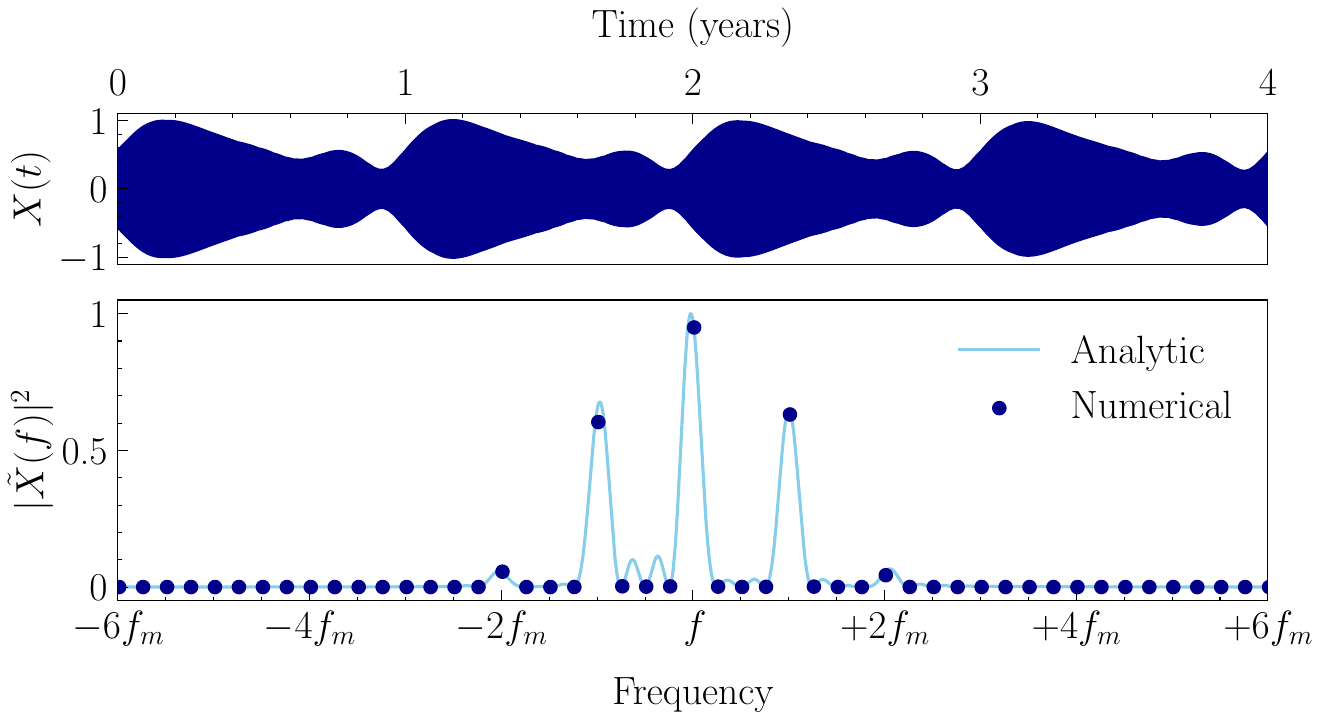}\label{fig:dmtf}
        %\\ (b) Waveform and spectrum of ULDM.
    \end{center}
    \caption{Waveforms and spectra of simulated GW~(upper) and ULDM~(lower) signals in $X$ channel.
    The dots in the lower panel of each subplot represent the spectrum obtained by Fourier transforming the corresponding time-domain signal shown in the upper panel, while the light-colored solid lines show the analytical spectra given by Eq.~(\ref{eq:X GW modul}) and Eq.~(\ref{eq:X DM modul}).
    The GW parameters are: frequency $f=1~\textrm{mHz}$, $A=1$, $\theta=\pi/4$, $\phi=\pi/3$, $\iota=\pi/4$, $\psi=0$, and $\varphi_0=0$.
    The ULDM signal is generated by a ULDM field with a Compton frequency $f_c = 1~\textrm{mHz}$ and amplitudes $a_x = -1$, $a_y = 1$, and $a_z =1+i$. 
    The signals are normalized such that the maximum of $X(t)$ is $1$. 
    }
    \label{fig:gwdmtf}
\end{figure*}

In Fig.~\ref{fig:gwdmtf}, we show the simulated GW and ULDM signals in $X$ channel for a set of source parameters. In the upper panels of both subplots, we show the signals in time domain, where the modulation appears as a varying amplitude envelope.
In the lower panels, we compare the numerical spectra $|\tilde{X}(f)|^2$ (dots), obtained by Fourier transforming the time-domain signals, with the analytical ones (solid lines) given by Eq.~(\ref{eq:X GW modul}) and Eq.~(\ref{eq:X DM modul}). Because the simulated signals contain a finite number of data points, the spectra exist only on discrete frequency points.
The close agreement between the analytic curves and numerical points across frequencies validates our analytical approach.

%------------------------------------------------------------
\section{Detector parameters} \label{ap:detector}
The one-sided power spectral density of the noise in $X$ channel is given by~\cite{babak2021lisasensitivitysnrcalculations}
\begin{equation}
   S_n(f)= 16\sin^{2}(2\pi fL) 
    \left[\left( 3+\cos(4\pi fL) \right)S_{\text{acc}} + S_{\text{oms}}\right],
\end{equation}
where $L$ is the detector's arm length, $S_{\text{oms}}$ the optical metrology system noise and $S_{\text{acc}}$ the test mass acceleration noise, which are adapted as
\begin{align}
    S_{\text{oms}}  &= \left(\frac{2\pi f}{c}s_{\text{oms}}\right)^{2} \left[1 + \left(\frac{2 \times 10^{-3}~\textrm{Hz}}{f} \right)^{4}\right]~\textrm{Hz}^{-1}, \nonumber \\
    S_{\text{acc}}  &= \left( \frac{s_{\text{acc}}}{2\pi fc} \right)^{2} \left[ 1+\left(\frac{0.4 \times 10^{-3}~\textrm{Hz}}{f} \right)^{2} \right] \nonumber \\ 
    &\times \left[ 1+ \left(\frac{f}{8 \times 10^{-3}~\textrm{Hz}} \right)^{4} \right]~\textrm{Hz}^{-1}.
\end{align}
For LISA~\cite{babak2021lisasensitivitysnrcalculations}, $s_{\text{oms}} = 15\times10^{-12}~\textrm{m}$, $s_{\text{acc}} = 3\times10^{-15}~\textrm{m}\cdot\textrm{s}^{-2}$, and $L = 2.5\times10^6~\textrm{km}$.

%------------------------------------------------------------
\bibliography{ref}% Produces the bibliography via BibTeX.
\end{document}